# Acoustically Transparent Alumina-based Cranial Implants Enhance Ultrasound Transmission Through a Combined Mechano-Acoustic Resonant Effect

Mario Ibrahin Gutierrez[a], Pathikumar Sellappan[b,c], Elias H. Penilla[b], Irais Poblete-Naredo[d*], Arturo Vera[e], Lorenzo Leija[e], Javier E. Garay[b]

[a]*CONAHCYT – Instituto Nacional de Rehabilitación Luis Guillermo Ibarra Ibarra, Subdirección de Investigación Tecnológica, División de Investigación en Ingeniería Médica, Calz. Mexico Xochimilco, Col. Arenal de Guadalupe – Tlalpan, Alcaldía Tlalpan, Mexico City, 14389, Mexico.*

[b]*Jacobs School of Engineering Mechanical and Aerospace Engineering, University of California San Diego, San Diego, CA 92161, USA.*

[c]*Rolls-Royce High Temperature Composites, Cypress, CA 90630, USA.*

[d]*Centro de Investigación y de Estudios Avanzados del Instituto Politécnico Nacional, Cinvestav-IPN, Department of Toxicology, Mexico City, 07360, Mexico.*

[e]*Centro de Investigación y de Estudios Avanzados del Instituto Politécnico Nacional, Cinvestav-IPN, Department of Electrical Engineering, Bioelectronics Section, Mexico City, 07360, Mexico.*

*Corresponding author: Irais Poblete Naredo
 E-mail address: ipoblete@cinvestav.mx (I.P.N.)

**Abstract**: Therapeutic ultrasound for brain stimulation has increased in the last years. This energy has shown promising results for treating Alzheimer's disease, Parkinson's disease, and traumatic brain injury, among other conditions. However, the application of ultrasound in the brain should trespass a natural but highly attenuating and distorting barrier, the cranium. Implantable ceramic materials can be used to replace part of the cranium as an alternate method to enhance ultrasound transmission. In this work, it is presented the acoustic characterization of alumina ceramic disks that can be employed as cranial implants for acoustic windows-to-the-brain. Alumina samples were prepared using current-activated pressure-assisted densification and were acoustically characterized. Acoustic impedance and attenuation of the samples were determined for different porosities. Additionally, measured and modeled acoustic fields are presented and analyzed in terms of the total ultrasound transmitted through the ceramics. Results indicate a resonant behavior in the alumina disks when the thickness corresponds to a half-wavelength of ultrasound; this resonance permits a total of 95.4% of ultrasound transmission; for thicknesses out of the resonant zone, transmission is 53.0%. Alumina proves to be an excellent medium for ultrasound transmission that, in conjunction with its mechanical and optical properties, can be useful for cranium replacement in mixed opto-acoustic applications.

# 1. Introduction

The use of ultrasound for the modulation of brain activity as well as the prevention, halt, and reversion of brain damage [1] is having promising results. In animal and human models, ultrasound acts as a neuromodulator of both excitatory and inhibitory neuronal pathways [2]. In diseased murine models, for instance, ultrasound enhances the release of neurotrophic factors in the brain [3] for the treatment of lipopolysaccharide and aluminum induced Alzheimer's disease models [4]. Experimental proposals using low intensity pulsed ultrasound (LIPUS) for the prevention and early treatment of ischemic stroke [5], vascular dementia [6], and traumatic brain injury [7] have also been presented. In humans, the neuromodulatory action of ultrasound was assessed in the motor and somatosensory cortex, visual cortex, and deep brain areas such as caudate nuclei and thalamus with both excitatory and inhibitory effects [2]. Moreover, the therapeutic potential of ultrasound has been explored for the treatment of mood and pain disorders [8,9], Alzheimer's disease [10], loss of consciousness after brain trauma [11], and cancer [12].

The application of ultrasound to the brain is limited by the cranium, which reflects about 50% of the waves due to impedance mismatch, and only transmits 3% to 30% of the energy [13]. Initially, it was proposed to deliver the energy directly to the brain using craniotomies [14], but the invasiveness of this procedure and the incapability to be used for multiple ultrasound applications had urged researchers to propose alternative methods [13]. When the ultrasound travels through the cranium, the transmitted waves are highly distorted by amplitude and phase aberrations [15], which complicates focalization at small target regions [16]. Despite that, transcranial ultrasound is being used in humans by combining a high number of mono-element piezoelectric transducers placed around the head [9,17]. For this, the therapy setup is customized according to the cranium thicknesses of each patient with the help of magnetic resonance imaging (MRI), and the stimulation is often applied simultaneously with MRI [8]. Although the solution is effective for certain single-application therapies, the technological requirements make this procedure expensive for protocols that would require repeated sessions during long periods of time.

Recently, implantable materials for cranium replacement have attracted interest as acoustic windows for therapy and diagnosis of brain pathologies [18,19]. Ceramics of 8-mol yttria stabilized zirconia (8YSZ) have been proposed as an acoustic window-to-the-brain (AWttB) that transmits up to 80.8% of the ultrasonic energy at certain thicknesses; this effect depends on a resonant behavior related to the ultrasound wavelength in the material [13]. The use of this specific material responded to a double intended use in which the window can be translucent for both optical and acoustical [13] energies for new opto-acoustic applications, besides the already presented in Optics [20]. Although 8YSZ transmits light at 1 mm thickness [21], it needs to be clarified if its transmission level is enough at the average human cranium thickness, which ranges from about 3 mm to 15 mm [22]. Furthermore, medium-term degradation of 8YSZ implants is still under discussion with inconclusive results [23].

More biocompatible materials with increased opto-acoustic transmission should be studied to improve both ultrasound and light transmission and the long-term stability of cranial implants. Aluminum oxide, $Al_2O_3$, or alumina, is a translucent ceramic material in a wide range of optical wavelengths and it is nearly transparent in the infrared zone [24]. Alumina has been used in a wide variety of fields, starting from militia, in transparent armor and radar wave absorption [25], to medicine, in implantable prosthesis [26]; these applications make use of the great properties of this material like the hardness, the scratch resistance, and the biocompatibility [26]. The optical properties of alumina are exploited as transparent windows [27] in devices like vacuum chambers for spectroscopy, watches, cellphone cameras, and lamps [28]. All these mechanical and optical properties of alumina, together with the high level of biocompatibility [26], make this material suitable to be used as an implantable ceramic for AWttB to improve ultrasound transmission for repeated brain therapy, which is the proposal of this paper.

## 2. Materials and Methods
### 2.1. Alumina sample preparation

Commercially available α-$Al_2O_3$ (99.99% purity, Taimei Chemicals, Japan) was densified by CAPAD (Current-Activated, Pressure-Assisted Densification) [27] to produce nine fully dense bulk disk-shaped specimens with 19 mm diameter limited by the size of the molds; this diameter does not have an impact in the results since ultrasound transmission is being studied through the sample thickness. The $Al_2O_3$ polycrystals were mechanically polished to their final thicknesses (1.23, 1.94, 3.80, 4.10, 5.15, 6.09 mm, and 3 samples with 5.60 mm). Samples of 4.5 mm average thickness with different densities of 59.9%, 64.0%, 69.9%, 74.9%, and 90.2% were made by adjusting the relationship between the processing temperature in CAPAD and the mass of the starting powder. For more details about sample preparation please consult supplementary material.

## 2.2. Acoustic characterization

The alumina samples were acoustically characterized with pulse-echo [29] using a 20 MHz transducer (V116-RM, Olympus Corp., USA) driven with a -100 V-pulse of 20 ns produced by a pulser (5073PR, Olympus Corp, USA). The ultrasound fields were measured with a wideband hydrophone (HNP-1000, Onda Corp., USA) by using the setup of Fig. 1(a); the samples were attached to the transducer radiating surface (7310, Mettler Electronics Corp., USA) using silicone glue (Mil'U, Cromher, Mexico). The transducer was driven with a signal generator (Array 3400, China) using 25 cycles 10 $V_{p-p}$ sine tone-burst, which produced an acoustic intensity of 70 mW/cm$^2$ and 46.8 kPa average acoustic pressure at the transducer radiating surface. The level of ultrasound transmission through the alumina was determined by averaging the pressure data after the samples that fell into an imaginary cylinder of 8 mm radius and 1 cm depth. This volume was chosen to prevent acoustic field interferences coming from the transducer surface that was not covered by the ceramic samples (see the full details in the supplementary material).

## 2.3. Modeling conditions

To deeply understand the factors involved in the ultrasound transmission, the system was modeled with the finite element method using COMSOL Multiphysics 5.5 (COMSOL Inc., Sweden). The geometry was simplified assuming the problem axisymmetric [13]. The problem was solved by assuming continuous wave and frequency dependent propagation. Ultrasound was produced by a vibrating planar piezoelectric disk of barium titanate (BaTiO$_3$) with 2.51 mm thickness and 2.1 cm radius with a frontal layer of glass of 160 μm thickness. Between the sample and the transducer, a 100 μm layer of silicone glue was included. The mechanical and acoustic properties of the materials used for the models are shown in Table 1S [30–33].

Two models were carried out by using the geometry shown in Fig. 1(b). For the first one, a structural mechanical simulation, it was assumed that the alumina disk vibrates freely to study the vibrational response of the samples. The ceramic disk thickness varied from 1.00 mm to 12.00 mm, and the radius from 5.00 mm to 15.00 mm at steps of 0.20 mm and 0.25 mm, respectively. Acceleration of the top surface was averaged for each parameter combination; normalization of the results was applied with the maximum average acceleration. The second set of simulations was carried out to calculate the total transmitted ultrasound through the sample. For this, the piezoelectric transducer was driven with a sine signal of 10 $V_{p\text{-}p}$ in accordance with the experiments. The piezoelectric disk vibrated at a stable frequency, transmitting the vibration through the other layers. The ultrasonic waves were produced by the normally accelerating surface of the piezoelectric disk in both directions, i.e. forwards to the layers of glass, glue, and alumina, and backward to the piezoelectric disk. At the thickness extensional (TE) mode of the alumina, the forward ultrasound wave will be reflected back at the alumina-water interface with the same phase of the main wave, which increases the resulting pressure amplitude; this process is repeated in to forward direction when the waves reach the back of the transducer. The equations that describe this behavior are those of piezoelectric materials fully detailed in the literature, with special reference to equations 4 to 7 in Ref. [13]; these domain equations and boundary conditions are detailed in the supplementary material.

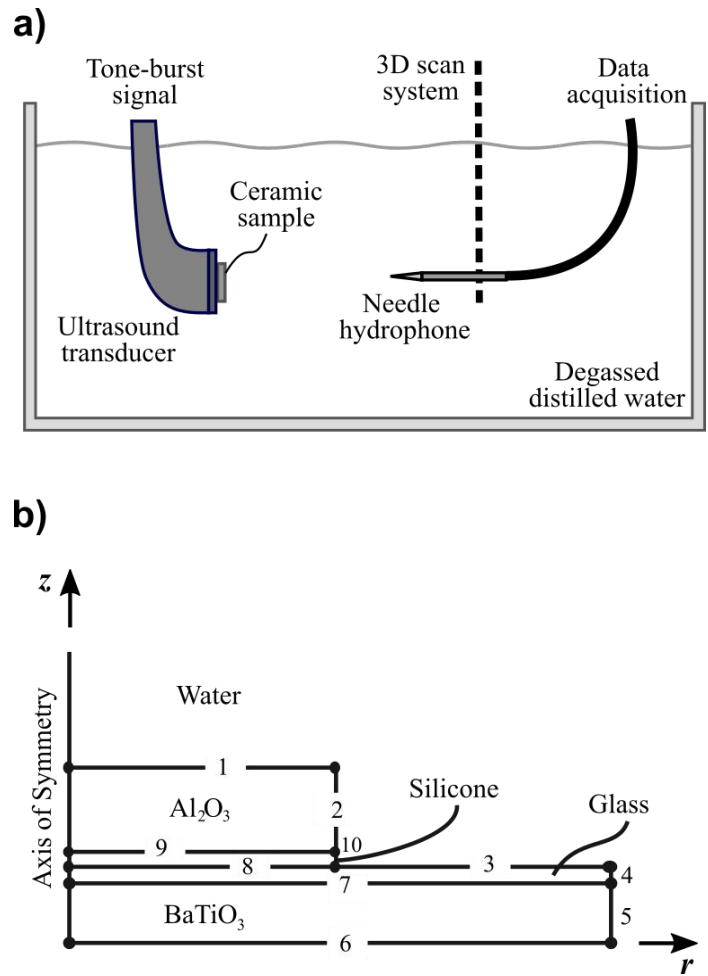

**Fig. 1.** (a) Experimental setup for measuring the acoustic field transmitted through the ceramic sample; acquired data were processed off-line. (b) Geometry for FEM analysis. External boundaries of water container were set at 8 cm radius and 20 cm depth, which are far enough to not interfere with the region of interest.

## 3. Results

### 3.1. Alumina characterization

The effect of CAPAD processing temperature in the relative density (see Eq. 1S) of alumina samples can be seen in Fig. 2(a). The relative density increased with temperature from 64% at 950 °C to 99% at 1200 °C. Then, the samples were characterized to determine the level of ultrasound transmission through them, either by itself (attenuation), or in relation with other media or tissues (impedance). Acoustic impedance and attenuation have a strong relationship with the porosity of the alumina, which is shown in Fig. 2(b, c), respectively. The acoustic impedance is inversely related to the porosity, both depending on density; conversely, the attenuation is directly related to the porosity.

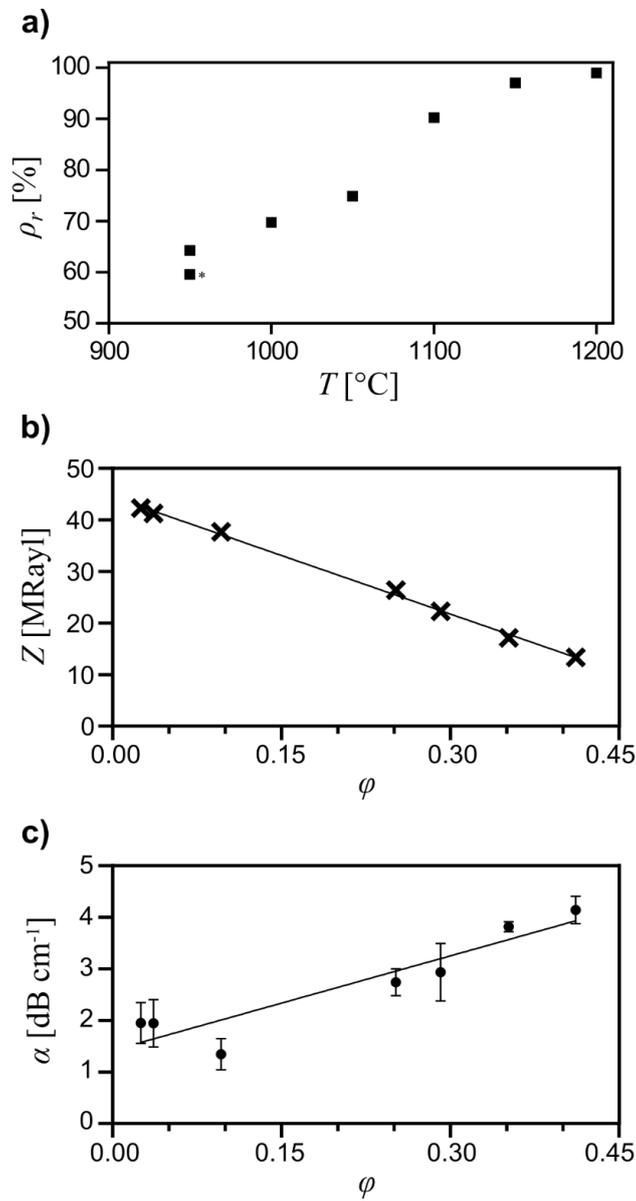

**Fig. 2.** Physical and acoustical characterization of the alumina ceramic disks. a) Relative density ($\rho_r$) of alumina disks for different CAPAD processing temperature ($T$) at 105 MPa. Data point marked with (*) corresponds to 950 °C and 35 MPa of pressure. b) Acoustic impedance ($Z$) and c) acoustic attenuation ($\alpha$) at 20 MHz (number of samples = 4) for different porosities ($\varphi$). Since some samples were thin, the dispersion of attenuation at certain zones is large.

### 3.2. Ultrasound transmission and mechano-acoustic resonance analysis

The acoustic field transmitted through different alumina disks was measured and modeled with FEM. Fig. 3(a) shows the modeled vibration distribution of the ceramic at the TE mode under free-vibration conditions. The normal acceleration distribution shows a main lobe at the central region of the disk, with a smaller variation close to the border. Fig. 3(b) shows the measured acoustic pressure at 2 mm at TE mode of a 5.4-mm-thick 10-mm-radius alumina disk (0.5$\lambda$ thickness) compared with the acoustic pressure pattern measured without the sample, i.e. at free-field condition. The modeled vibration at TE mode is congruent with the pattern of the measured field after the sample (Fig. 3(b), solid line), but it is different from the pattern of the free-field condition (Fig. 3(b), dashed line). This transmission was significantly increased at the TE vibration mode of the disks, probably because this resonant effect overcomes other reductions of transmission, for instance, those produced by impedance mismatching.

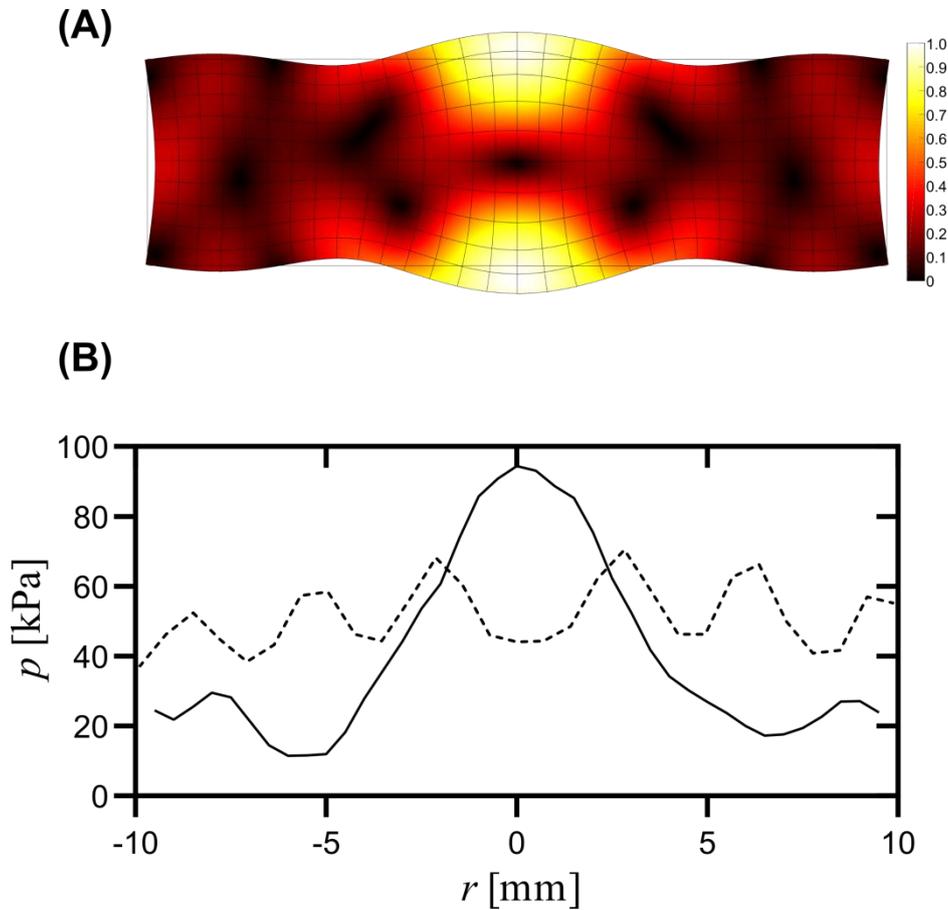

**Fig. 3.** Free-vibration and measured transmitted field through the 0.5$\lambda$-thickness alumina disk. a) Modeled normal acceleration of alumina during free vibration (normalized linear color scale). b) Measured peak-to-peak acoustic pressure at 2 mm from the sample (*solid line*) and at 2 mm from the transducer without the sample, i.e. free-field (*dashed line*).

Furthermore, because the measured pattern after the ceramic is different than the pattern measured at free-field condition, we could assume ceramics do not only transmit the ultrasound, but they also modify the field. Moreover, because of the measured acoustic pattern is congruent with the modeled distribution of Fig. 3(a), the produced field after the sample would be influenced by the self-vibration of the alumina disk. As shown later in Fig. 5, the average pressure after the sample adequately corresponds to that obtained in the experiments.

These simulations under free-vibration conditions also permitted the study of the effect of the radius and thickness in the amplitude of the acceleration of the disk surface; for determining this, both parameters were varied in the model. Figure 4 shows the resonances at the TE mode for $\lambda/2$ (being theoretical $\lambda$ =

10.9 mm at 1 MHz for full dense alumina), and at the second harmonic for $\lambda$. It can be observed that for small radii, radial and TE modes are coupled, which modifies the vibration patterns and should increase the complexity of the calculations. However, since this application is only intended for TE mode, it is suggested to use only ceramics at a radius larger than 8 mm where the effects of other vibration modes are negligible. This would permit stable and predictable vibration patterns.

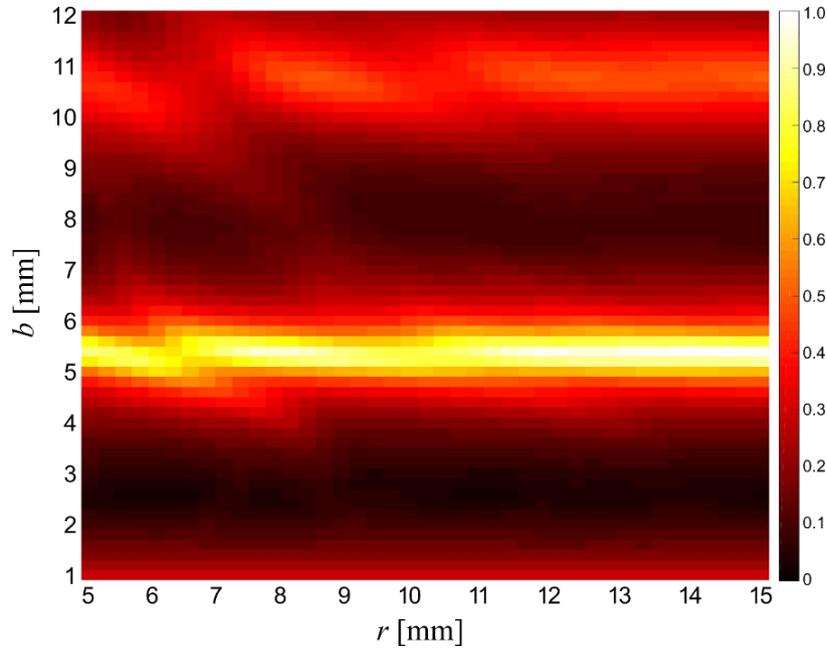

**Fig. 4.** Modeled normalized average acceleration at the alumina surface for different thickness and radius combinations under free-vibration conditions Experimentally, half-wavelength thickness was 5.6 mm due to differences in acoustic properties of single samples.

Based on the measured and modeled 2D fields, the quantitative average transmitted ultrasound for different thicknesses can be seen in Fig. 5. For the model, the efficiency of electrical-to-mechanical energy conversion was set to 61% as measured for this transducer at 1 W. From this figure, the ultrasound is efficiently transmitted when the thickness corresponds to multiples of the half-wavelength of ultrasound in the material. The maximum transmission of 95.4% of the reference pressure in free-field condition (with no sample attached at the transducer) was obtained at $0.5\lambda$, according to the theory. However, the transmission at other regions is about 53% of the reference pressure, which is still large enough to be considered a better ultrasound transmission than the cranium, which is about 2.8% at 1 MHz and 6 mm thickness measured [13].

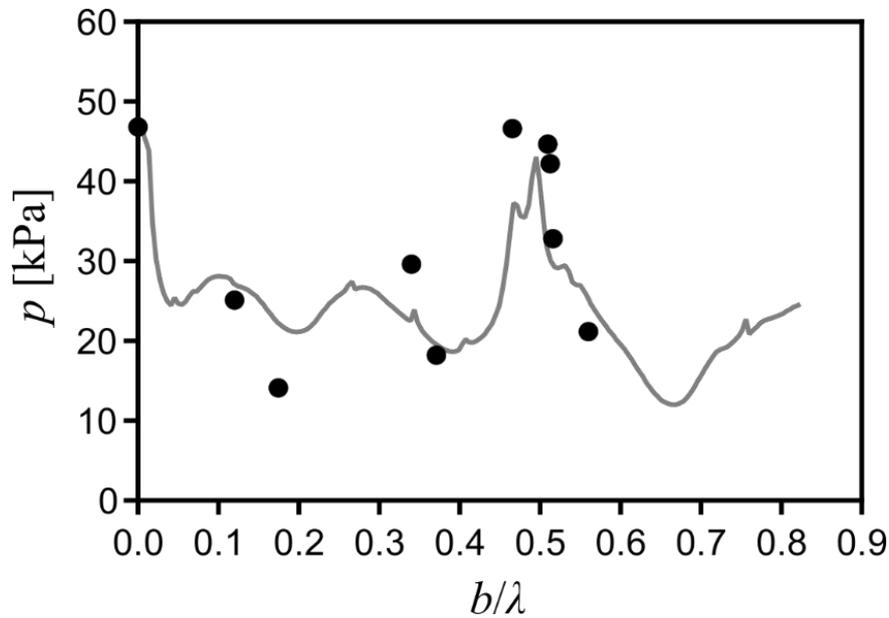

**Fig. 5.** Experimental (dot) and modeled (line) average acoustic pressure through alumina for different thicknesses (*b*). Ultrasound transmission was improved when the thickness was about 0.5$\lambda$ due to a resonant acoustic effect in the ceramic. The average transmitted pressure was about 53% of the maximum (without the ceramic at free field condition) for no-resonant thicknesses.

## 4. Discussion

The acoustic field transmitted through various alumina disks was measured and modeled with FEM. The acoustic transmission was averaged and quantified after the disks and was compared with the free-field condition (no-sample). It was determined that ultrasound transmission, under our experimental conditions, depends on the alumina thickness and the ultrasound wavelength, with a significant increase at the TE vibration mode. Moreover, the experimental radiating patterns indicate a secondary induced vibration in the samples that modifies the produced acoustic field. The measured field after the sample was composed of the field produced by the transducer and the secondary field produced by the ceramic's vibration.

The porosity of the samples was increased by reducing the CAPAD processing temperature. These samples were acoustically characterized to determine the effect of porosity on the acoustic properties (see the supplementary material). Acoustic impedance and attenuation exhibit strong dependence on the porosity of the alumina, as shown in Fig. 2(b, c), respectively. The acoustic impedance has a negative linear trend versus porosity while the opposite is true for the attenuation, showing a positive correlation with porosity. For full dense samples, the attenuation is smaller, being 1.9 dB cm$^{-1}$ @ 20 MHz (0.095 dB

cm$^{-1}$ MHz$^{-1}$), but the impedance is relatively large, 42.3 MRayls. Conversely, for 59.9% dense samples, attenuation increases by a factor of two (4.1 dB cm$^{-1}$), and the impedance decays to 13.4 MRayls. These values are smaller than those reported for 8YSZ [13], which is positive for ultrasound propagation. Moreover, alumina provides an extra advantage due to its improved mechanical properties [25,26,28].

To determine the natural vibration of the alumina disks, it was carried out an initial stage FEM model including a mechanical input, as depicted in Fig. 1(b). This analysis under free vibration conditions produced the 2D image of Fig. 4, which shows the main resonance at the TE mode when $\lambda/2$ and the second harmonic at $\lambda$. The location of TE mode is more stable for radii larger than 8 mm (diameter about 3 times the thickness at TE mode), with radial vibrations interfering the smaller disks. At the second harmonic, radial modes are still present for larger diameters. When working at the main TE mode, radial interference is significantly reduced after 10 mm, which permitted to have more defined TE vibration patterns. For the intended application of this material, radial vibrations are not of our interest since the main propagation would be in TE direction. However, having coupled vibration modes would complicate the produced acoustic fields because of the modulation with radial displacements of the TE vibration.

Fig. 3(a) shows the modeled vibration distribution of the ceramic at the TE mode under free vibration conditions. The normal acceleration distribution shows a main lobe at the central region of the disk, with a smaller variation close to the border. Fig. 3(b) shows the measured acoustic pressure at 2 mm from the 0.5$\lambda$ alumina that adequately corresponds to the modeled free-vibration at the same thickness. The field pattern after the ceramic is different than the pattern measured at free-field conditions (Fig. 1S). This observation suggests that the ceramic not only transmits ultrasound but also alters the field. However, the presence of an acoustic pattern consistent with the vibration distribution obtained from models under free-vibration conditions may indicate that the resulting field after passing through the sample is primarily influenced by the mechanical self-vibration of the alumina disk. The same effect can be seen in Fig. 1S(d), where the measured field distribution at alumina resonance has a dominant contribution of the free vibration pattern of Fig. 3(a), while its modeled counterpart is composed of a mixed pattern with a predominant piston-like behavior [13].

The transmitted average pressure after the ceramic for the model and the measurements is shown in Fig. 5. These measurements were warried out at 70 mW/cm$^2$ of acoustic intensity, which is far from producing non-linear effects [34,35]. At the resonant thickness, the transmission was about 95.4% of the measured pressure in free-field condition (with no sample attached at the transducer). This should be contrasted with the measured transmission in the cranial the bone which is about 2.8% at 1 MHz and 6 mm thickness [13]; other groups have reported transmissions of 4% at 1.8 MHz and 4.44 mm thickness [33], and 31% at 500 kHz and 4.3 mm thickness [22]. Moreover, these values also showed an improvement in the transmission level through alumina compared to the levels using 8YSZ disks in those same regions [13]. The ringing effect, or an oscillation-like behavior of the transmitted pressure versus thickness at the region before the resonance, was produced by constructive and destructive interference in the disks. However, the average transmitted ultrasound at that region (53%) is still considerable compared with bone and other ceramic materials proposed for this application. After the resonant zone, the transmission was importantly reduced to about 27% due to wave cancellations into the ceramics that provoke anti-resonant effects at those thicknesses.

The use of alumina cranial implants could be an attractive alternative for cranioplasties, or when there is no other alternative for long-term ultrasound therapy. Other groups have proposed non-invasive high intensity ultrasound techniques with multiple transducers for focusing in order to reach the interior parts of the head [17]. Although their advantage of being non-invasive, these technologies require prolonged time for therapy planning and expensive sessions when using real-time imaging technologies for monitoring [8,17]. This makes the therapy time-consuming and expensive when repeated sessions of low intensity ultrasound, as LIPUS, are required, specifically when non-thermal effects are desirable [13]. Using an acoustically transparent implant as cranium replacement will broad the ultrasound applications in brain, with potentially mixed opto-acoustic capabilities. Moreover, other applications could arise with the incorporation of implanted piezoelectric nanoparticles [36]. These particles, when being excited by ultrasound, can produce voltage that will electrically stimulate the brain regions where they are implanted. This stimulation combined with ultrasound would have potential benefits in regenerative medicine, neuromodulation, and cancer [37].

## 5. Conclusions

The use of alumina implants to replace a part of the cranium is a convenient alternative to metallic and polymeric materials. Alumina is stiffer than both of these metals and polymers, with midway, but adjustable, density. Although increasing the porosity would have a certain detrimental effect on the mechanical properties of the ceramic, it would provide a mechanism to adjust the acoustic properties in accordance with the patient's requirements (cranium thickness). Acoustic absorption of this material would provide a relatively small heat loss, even for the largest porosities analyzed. Acoustic impedance can be adapted to the necessities, specifically to the required final thickness of the cranium. With these characteristics, and our findings of resonant behaviors at half-wavelengths, it would be possible to adapt this material to different practical situations where therapeutic ultrasound is required to be applied in the brain. The 95.4% ultrasound transmission at the resonant thickness in alumina represents an improvement of transmissions through bone (4.4%) and through 8YSZ (80.8%). The biocompatibility of alumina reported by other groups has encouraged us to propose this material as a cranium replacement [26]; however, the biocompatibility of our samples should be verified during the next stage of this research, i.e. animal tests. Although further research is needed to ultimately achieve fully implantable alumina, based on the results presented in this study, we conclude that alumina provides excellent ultrasound transmission and can be used as an AWttB.

**Declaration of Competing Interest**

The authors declare that they have no known competing financial interests or personal relationships that could have appeared to influence the work reported in this paper.

**Acknowledgements**

The authors would like to thank Rubén Pérez Valladares for his technical support during samples characterization. This work was supported by CONAHCYT [grant number 257966]; CYTED-DITECROD [grant number 218RT0545]; and AMEXCID-AUCI 2018-2020 [project number IV-8]. JEG gratefully acknowledges the financial support of the Office of Naval Research (ONR).## References